# Microwave programmable response of Co-based microwire polymer composites through wire microstructure and arrangement optimization

A. Uddin[1], F.X. Qin[1*], D. Estevez[1], S.D, Jiang[2], S.A. Jawed[3], L.V. Panina[4,5] and H.X. Peng[1]

[1]*Institute for Composites Science Innovation (InCSI), School of Materials Science and Engineering, Zhejiang University, 38 Zheda Road, Hangzhou, 310027, PR. China*
[2]*School of Materials Science and Engineering, Harbin Institute of Technology, Harbin, PR. China*
*PR. China*
[3] *Karachi Institute of Economics and Technology, Karachi, Pakistan*
[4]*National University of Science and Technology MISIS, Institute of Nanotechnology and Novel Materials, Moscow 119049, Russia*
[5]*Institute of Physics, Mathematics & IT, Immanuel Kant Baltic Federal University, A. Nevskogo 14, 236041 Kaliningrad, Russia*

**Abstract**

Traditional approaches to realize microwave tunability in microwire polymer composites which mainly rely on topological factors, magnetic field/stress stimuli, and hybridization are burdensome and restricted to rather narrow band frequencies. This work presents a facile strategy based on a single component tunable medium to program the transmission response over wide frequency bands. Structural modification of one type of microwire through suitable current annealing and arrangement of the annealed wires in multiple combinations were sufficient to distinctly red-shift the transmission dip frequency of the composites. Such one wire control-strategy endorsed a programmable multivariable system grounded on the variations in both the overall array conductivity or "effective" area determined by the wires arrangement and the relaxation time dictated by the annealing degree of microwires. These results can be used to prescribe transmission frequency bands of desired features via diverse microwire arrays and microwave performance from a single component to composite system design.

---

[*] corresponding author: faxiangqin@zju.edu.cn

# 1. Introduction

Recently, polymer-based composites incorporating arrays of ferromagnetic glass-coated microwires have gained much attention due to their Giant Magneto-impedance (GMI) effect and tunable soft magnetic properties which confer them potential applications in electronic sensing, electromagnetic shielding, microwave absorption and structural health monitoring [1,2]. Although electromagnetic functionalities (e.g., metamaterial wave phenomena/microwave absorption) have been shown in composites containing such wires, their electromagnetic response is limited to topological factors and magnetic field/stress stimuli [3–7]. Most recently, inspired by the enhancement in the mechanical and electromagnetic properties of hybrid composites by the addition of nano-carbons [8–11] we adopted the multiscale design philosophy by adding carbon fillers into the microwire composites [12,13]. However, such an approach required delicate control and fine-tuning of the constitutive parameters of the nano-carbons.

Considering that the wires' intrinsic electromagnetic parameters relate to their microstructure, we propose to add to these methods a programming-based strategy by incorporating arrays of the same type of microwire but with different internal structure and adjusting their combination within the composite. Apart from doping with metallic elements such as Cu, Cr, Nb, the microstructure of the microwires and their electromagnetic properties can be conveniently tailored by current annealing, which promotes internal stress relaxation or redistribution and alteration in domain structure. Thus, the approach presented here is to implement different DC current-annealing conditions on Co-based wires of the same composition and tune their transmission parameters through distinct dynamic wire-wire interactions, which is enabled by the arrangement of structurally different wires within the microwire array composite. Both transmission magnitude and transmission dip frequency proved to be tunable with the systematic combination of the as-cast and annealed wires. Our single component-control strategy provides an accessible platform for programming and prescribing the ultimate electromagnetic properties of microwire array composites for a wide frequency range, which is hard to design and often requires plenty of components to fulfill requirements for each particular application.

## 2. Experimental Method

### 2.1. Current Annealing of $Co_{60}Fe_{15}Si_{10}B_{15}$ glass-coated microwires

$Co_{60}Fe_{15}Si_{10}B_{15}$ glass-coated microwires with total wire diameter $D_w$=35.4 µm and metallic core diameter of $D_m$=27.2 µm were selected for our study. The microwires were fabricated by a modified Taylor-Ulitovsky method [14] by simultaneously drawing and quenching the molten master alloy. 5 grams of the master alloy with the selected composition were put into a Pyrex glass tube and placed within a high-frequency inductor heater. While the metal melted, the portion of the glass tube adjacent to the melting metal softened; enveloping the metal droplet. A glass capillary was then drawn from the softened glass portion and wound on a rotating coil. By adjusting the drawing conditions, the molten metal filled the glass capillary and the microwire was formed in which the metal core was completely coated by the glass shell. To modify the structural features of the microwires and therefore, electromagnetic properties, as-cast wires (labeled as A) of about 12 cm in length were annealed at different DC currents of 20 mA, 30 mA and 40 mA for 10 minutes. Prior to current-annealing, the glass coating of the wire was mechanically removed from both ends to allow electrical contact (Fig. 1). In addition, the DC resistance was monitored after each current application with a digital multimeter showing a considerable drop for the wire treated at 40 mA, which indicates a possible onset of crystallization. Such drop is related to large grain size due to the rise in temperature caused by the current that will lead to less electron scattering at the reduced grain boundaries [15]. The crystallization starts initially at the surface of the wire and grows slowly with islands of crystallites nucleating homogeneously all over the wire (as will be elucidated in Section 3.2). This leads to bigger grain size formation and lower resistance as it crystallizes. After current-annealing, the morphology and topography of the wires were studied by scanning electron microscopy (SEM) in a Hitachi S-4800 cold field emission scanning electron microscope and a Bruker icon atomic force microscope (AFM). Thermal, structural and magnetic properties of the microwires were evaluated by Differential Scanning Calorimetry (DSC 204 HP Phoenix, the heating rate of 10K/min), X-Ray Diffraction (BEDE multi-function high-resolution X-ray diffractometer) and Quantum Design PPMS-VSM at room temperature, respectively.

[Figure 1 here]

2.2. *Preparation and electromagnetic characterization of $Co_{60}Fe_{15}Si_{10}B_{15}$ glass-coated microwire/ silicon resin composites*

The as-cast and annealed wires were incorporated into silicon-resin matrix in different arrangements and combinations (Fig. 2). The initial arrays were composed of the same type of microwire, either as-cast A (0 mA), annealed wires B (30 mA) or C (40 mA), Fig. 2 a. Subsequently, the as-cast wires were combined with B annealed wire (Fig. 2 b) and C annealed wire (Fig. 2 c) placing the wires in different sequences. The composites were prepared by aligning six wires in a parallel manner between two facing molds with a fixed wire spacing of 2 mm. Samples with dimensions of 22.86 x 10.16 x 2 mm$^3$ were realized after curing at 125 °C for 20 min the silicone resin/curing agent mixture. Scattering S-parameters of the composites were measured with Rohde & Schwarz ZNB 40 vector network analyzer (VNA) by using a WR-90 waveguide in $TE_{10}$ dominant mode from 8.2 to 12.4 GHz. Before the measurements, the VNA was calibrated by the TRL (thru-reflect-line) calibration method [16].

[Figure 2 here]

### 3. Results & Discussions

*3.1. Structural and thermal properties of $Co_{60}Fe_{15}Si_{10}B_{15}$ glass-coated microwires*

In order to check the effect of annealing treatment on the structural and thermal properties of the microwires, we explored DSC and XRD analyses. The increase in temperature during current annealing depends not only on the current density but also on the microwire diameter [17]. To evaluate the temperature reached due to current annealing, the following equation of conversion of electrical-to-thermal energy can be used [17]:

$$j^2 \rho = \frac{4 D_w}{D_m} \left( h \left( T - T_{ex} \right) + \varepsilon \propto \left( T^4 - T_{ex}^4 \right) \right) \qquad (1)$$

where ρ is the electric resistivity, $h$ is the coefficient of Newton's law of convection cooling, ε is the emissivity coefficient, α=5.67·10$^{-8}$W/m$^2$/K$^4$ is the Stefan-Boltzmann constant and $T_{ex}$ is the ambient temperature. In the case of microwires with a metallic core diameter in the range of 20 microns, currents of 25-50 mA correspond to temperatures of 400-550 K which are high enough for structural relaxation and partial relief of internal stresses [18], but lower than the crystallization temperature, $T_x$ of 720 K shown in Fig. 3 a. DSC curves of the as-cast wires and 30mA and 40mA annealed wires reveal the influence of current annealing on the structural relaxation and crystallization processes. For the as-cast wire, a very small Curie peak, around $T_c$ of 635 K reflects a large concentration of internal stresses [19]. In addition, a two-step crystallization process is also noticed. Annealing the wire at 30 and 40mA results in a splitting of the $T_c$ into two peaks confirming the presence of two magnetic phases and appearance of small clusters with different local environment [20]. The higher intensity of these peaks and increase of $T_c$ are related to a reduction in internal stresses and structural relaxation processes [19,20]. In addition, the broadening of the crystallization peak after annealing and increase of $T_x$ with respect to the as-cast wire, demonstrate the larger fraction of nanocrystals (area under this peak is proportional to the fraction of nanocrystals) and stress relief in the annealed samples [19].

The XRD spectra further confirm that the microwire maintains its amorphous structure when annealed at the maximum current intensity of 40 mA (Fig. 3 b). Therefore, annealing of the microwires mainly induced stress relief, structural relaxation, and homogenization. However, these subtle structural changes in the microwires will have a considerable influence on the magnetic and electromagnetic properties of arrays composites, as will be elucidated in the following sections.

[Figure 3 here]

Figure 4 shows the high-resolution transmission electron microscopy (HRTEM) images for the as-cast and current annealed wires i.e., 30 mA and 40 mA respectively. For the as-cast wire (Fig. 4 a), the microstructure consists of nanocrystalline droplets of about 25 nm embedded in the amorphous matrix exhibiting lighter and darker contrast due to the different composition of the two separated phases [21]. The inset indicates the selected-area electron diffraction (SAED) pattern which confirms the nanocrystalline nature of the as-cast sample not evidenced from the XRD results showing nanocrystalline spots embedded in the predominant amorphous phases. The droplets are also seen for the annealed samples along with very fine particles smaller than 5 nm precipitated uniformly within the amorphous matrix for the 40 mA annealed-sample which confirms the trend seen for the $T_c$ (Section 3.1). Because the increase in the atomic number of the constituent elements decreases the intensity of the transmitting electrons, we deduced that the darker droplet-like nanoparticles mainly consist of the heavier elements Co and Fe, while the brighter droplets correspond to the lighter elements Si and B. The bright outline around the droplet in the inset of Fig. 4 b (pointed at by yellow arrows) also supports this interpretation, i.e., the presence of Si, B-enriched region having high diffusivities [21]. The phase separation is also evidenced by the high-angle annular dark-field scanning transmission electron microscopy (HAADF-STEM) image of the as-cast sample (Fig. 5 a) which shows bright spots indicating the presence of segregation or inhomogeneous regions in which relatively heavy elements are localized (in the HRTEM images, the relation of the image contrast was opposite). From the 2-D energy-dispersive spectra (EDS) maps, Fig 5 (a1-a3), these regions correspond mainly to Co and Fe, as expected.

[Figure 4 here]

[Figure 5 here]

*3.2. Magnetic and magneto-impedance properties of $Co_{60}Fe_{15}Si_{10}B_{15}$ glass-coated microwires*

Figure 6 shows the magnetic hysteresis loops (Fig. 6 a) and the magnetic field dependence of the magneto-impedance ratio for the as-cast and annealed samples annealed with different currents (Fig. 6 b). Both the magnetization process and magneto-impedance ratio of microwires are defined by the magnetic anisotropy affected e.g. by current annealing [18,22]. Large internal stresses give rise to a large anisotropy which is evidenced in the M-H loop of the as-cast wire (Fig. 6 a). Annealing the microwires progressively induce anisotropies along local magnetizations (slope of the loops gradually decreases) due to short-range pair ordering and a circular magnetic field generated by the current which aligns the easy anisotropy axes. The slightly larger anisotropy of 40 mA sample compared to that of 30 mA sample (inset Fig. 6 a) may be due to the closeness of crystallization, as mentioned previously. Current annealing also improves the magnetic softness evidenced by the drastic decrease in coercivity from 15 Oe for the as-cast wire to 0.8 Oe for the wire annealed at 40 mA due to stress and structural relaxation during the annealing process. From the magneto-impedance plot at 200 MHz (Fig. 6 b), the as-cast wire is characterized by a slightly distorted single peak which evidences a system with the easy anisotropy parallel to the high-frequency current [23]. This is in disagreement with the M-H loop of the as-cast wire in which an inclined shape is indicative of the presence of circumferential anisotropy. To explain such discrepancy, we first should consider the differences between these two measurements. M-H curves correspond to static magnetization and thus bulk magnetic properties can be obtained while magnetoimpedance measurements are of dynamic character in which magnetization processes depend on ac current and frequency. Second, we should consider the magnetic structure of these type of wires in which both circumferentially magnetized shell and longitudinally magnetized inner core exist [24]. Both conductivity channels are subjected to the frequency dispersion due to the ac magnetic susceptibilities and their interplay especially near the ferromagnetic resonance results in complex transformations of MI when changing the frequency from MHz to GHz. In previously published reports, only the influence of the shell is taken into account to explain the impedance field dependence of these wires, but thanks to our recently developed broadband impedance measurements (which will be reported in next works) we are able to disclose the role of the inner core as well. At 200 MHz, the single peak is indicative that the main contribution to the magnetization comes from the longitudinally magnetized inner core. At such frequency, the ac excitation current will penetrate deeper into the wire. Another evidence of the contribution of the longitudinally magnetized core is delivered by the dispersion curves of magnetoimpedance measured in a wide range, as explained in the Supporting Information (Section 1, Fig. S1 b). With increasing the frequency, the single peak transforms into asymmetrical double peaks (Fig. S1 a) demonstrating the gradual contribution to the magnetization of the circular domains at the wire surface shell which will be reflected in the shape of the M-H loops.

Annealing the wire originates a double peak asymmetrical magneto-impedance response due to a slight deflection of the longitudinal anisotropy. It should be noted that such improvement in magnetic softness does not benefit GMI; this can be ascribed to the increase in surface defects and roughness, as evidenced by the morphological and topological characterization of the microwires (Fig. S2 and Fig. S3). It causes fluctuation of local magnetic anisotropy at the surface and immobilization of surface spins [24–27].

[Figure 6 here]

Anyhow, the general magnetic behavior of the microwires depends intimately on the domain structure, which is determined by minimization of the magnetoelastic energy $K_{me}$ given by:

$$K_{me} = \frac{3}{2}\lambda_s \sigma_{ii} \qquad (2)$$

where $\lambda_s$ refers to magnetostriction coefficient, $\sigma_{ii}$ refers to the dominant internal stress component [28]. By adjusting the microstructure of the wires through annealing treatment, the internal stresses $\sigma_{ii}$ of the wires can be relieved or redistributed. The internal stresses can be expressed as:

$$\sigma_{ii} = \frac{2}{3}\left(\frac{\mu_0 M_s}{\lambda_s}\right)(H_{kfin} - H_{kin}) \qquad (3)$$

where the term ($H_{kfin}-H_{kin}$) refers to the decrease in anisotropy field due to the annealing treatment, $\mu_0$ represents the vacuum permeability and $M_s$ the saturation magnetization. Such internal stresses are introduced while fabricating the wires coupled with magnetostriction and are distributed in a complex way forming a unique domain structure [29]. From the above equation decreasing the anisotropy field results in a decrease in internal stresses. Such a decrease in internal stresses [30] reduces the magnetoelastic energy which modifies the domain structure [31] by inducing relatively well-defined magnetic domains in the 40 mA annealed wire (inset Fig. 6 a). The changes in magnetic properties of the wires after annealing will have an important influence on the electromagnetic properties of the composites, as will be elucidated in the next section.

*3.3. Transmission spectra of composites incorporating as-cast and current annealed wires*

In this section, we will show that by incorporating microwires with different structural features arranged in different combinations into a polymer matrix the electromagnetic parameters are readily programmable, in particular, the transmission coefficient, $S_{21}/S_{12}$ being a measure for the transmission loss is crucial in understanding the microwave properties of the composite samples. Fig 7 a shows the transmission spectra $S_{21}$ of the composites containing the individual as-cast wires (0 mA) and annealed wires (20, 30 and 40 mA). For the as-cast wire sample, a transmission dip occurs at 11.96 GHz corresponding to the Lorentz-type dielectric resonance or dipolar behavior of the short-cut wires [6]. Short-cut wire inclusions act as dielectric dipoles when interacting with the electrical component of waves. The dipole resonance can be written as $f_{res} = \frac{c}{2l\sqrt{\varepsilon_m}}$ (below the percolation threshold), where $c$ is 3 x 10$^8$ m/s represents the speed of light, $\varepsilon_m$ and $l$ denote the permittivity of the silicone matrix and microwire length, respectively [6]. Taking $\varepsilon_m$ as 2 and $l$ as 10.16 mm (fixed in our case) into the above equation, we obtain $f_{res}$ = 10.44 GHz, which is somehow lower but close to that of the identified dip in Fig. 7 a for as-cast wire. Increasing applied current on the wire shifts the transmission dip toward lower frequencies (red-shift). Initially, with a small amount of DC current, i.e. below 20 mA, $f_{res}$ is barely affected, whereas for currents above 20 mA, the red-shift is more evident. As we demonstrated in Sections 3.1 and 3.2, gradually annealing the microwires results in a relief of internal stresses and induced structural relaxation, therefore, the shift in the transmission-dip frequency is mainly caused by the consecutive structural changes experienced during the annealing process.

For the composites containing combinations of the as-cast and annealed microwires (Fig. 7 b and Fig. 7 c), the arrangement of the microwires has a profound impact on the transmission amplitude and transmission dip frequency of the composites. When different types of microwires are added in the same amount, i.e. AAABBB and AAACCC (Fig. 7 b), the transmission dip or dipolar resonance trends to lower frequency as the 30 mA-annealed microwire B is replaced by the 40 mA-annealed microwire C. This trend coincides with that of the composites containing the same type of microwire (AAAAAA; BBBBBB and CCCCCC), but with a most pronounced red-shift for the AAACCC sample.

When the wires are arranged in an alternate manner i.e. ABABAB and ACACAC, the transmission dip follows the same trend with frequency, i.e., red-shifting with the incorporation of 40 mA-annealed wire C (Fig. 7 a). Finally, when the middle microwires in the array ABABAB and ACACAC are switched, i.e. ABBAAB and ACCAAC, the critical frequency once again shifts on the side of the 40 mA-annealed wire C. In addition, the transmission amplitude of the composites also decreases as wire C is incorporated in the array (Fig. 7 a-c). Therefore, one can infer a dominant influence of the annealed microwires over the as-cast microwires on dictating the electromagnetic response of the composites.

[Figure 7 here]

The variations in transmission spectra feature in the composites could be explained by the dynamic wire-wire interactions considering the differences in domain structure and magnetic tensors between the as-cast and annealed wires incorporated in the array [30,32]. In this case, the dynamic magnetic interactions resulted from the coupling with an electrical component of incident waves rather than the magneto-static coupling, since 2 mm spacing is still too wide to induce meaningful magneto-elastic energy [28]. In the first case AAAXXX (where X corresponds to either 30 or 40 mA annealed wires), we can consider the largest "effective" area composed of equally structural microwires (three as-cast wires plus three-annealed wires) with different conductivity, larger for the annealed wires (Fig. 8 a). The same type of wires in a parallel manner constitute an increase of the total cross-sectional area of the wires, resulting in a diminution of the effective resistivity and hence a stronger skin effect [33]. Such "effective" area is decreased when varying the order of the wires in the arrangement, three wires of the same type are reduced to two (AXXAAX, Fig. 8 b) and one (AXAXAX, Fig. 8 c), therefore affecting the equivalent conductivity and shifting the transmission dip frequency of the composites.

[Figure 8 here]

In addition to structural relaxation, annealing also reduces the internal damping or relaxation rate $\Gamma$ on nanostructures which is inversely proportional to the electron relaxation time $\tau$, i.e., $\Gamma = h\tau^{-1}$, where $h$ is Planck's constant [34]. The increase in relaxation time with annealing might be related to the progressive switching of the transmission dip to low frequency for the arrays containing the annealed samples (Fig. 7). In fact, the decrease in resistivity for the 40 mA annealed-sample is due to the enhancement in nanocrystallinity (Fig. 4 c), carrier concentration and carrier mobility [35,36] therefore decreasing the internal damping.

Let us summarize our programming strategy in Fig. 9, in which a horizontal line represents the transmission dip frequency for the different arrays containing the as-cast A, annealed wires X and their combinations. The initial frequency band, i.e. the frequency ranges from the frequency of the composites containing merely as-cast wires *fA* to the frequency of the composites containing merely annealed wires *fX* can be broaden or shorten according to the annealing degree of the wires, which will modify *fX*. Random combinations of as-cast and annealed wires will locate the transmission dip frequency between the critical band determined by *fA* and *fX*. Finally, using the same amount of as-cast and annealed wires in a consecutive manner, the frequency falls out of *fA-fX* range, red-shifting progressively with the current annealing intensity. Our approach shows that incorporating wires having diverse structural features arranged in different combinations, the electromagnetic properties could be largely anticipated and programmed. Moreover, stress-relaxation and structural change of the wires through low-current annealing without varying the general nanocrystalline structure and without crystalline degradation are only necessary to induce a significant change in the signature of the transmission spectra. Because of the final electromagnetic properties are decided not only by the structural differences of the wires but also by the dynamic wire-wire interaction, changing the distance between the wire arrays could offer further tunability of the microwave response. Relevant work is now underway.

[Figure 9 here]

## 4. Conclusion

We proposed a strategy to program and formulate the microwave properties of microwire array composites via a single wire component subjected to different current annealing conditions and arrangements in the polymer matrix. Current-annealing the wire resulted mainly in stress and structural relaxation which was reflected in the change of Curie temperature, anisotropy field, domain structure and transmission responses of the microwires and their composites, respectively. The transmission dip frequency and magnitude of the composites containing either the same type of wires or combinations were mainly influenced by the incorporation of the wire treated at the largest current intensity. By altering the largest "effective" area which is composed by the equal combination of as-cast and annealed wires in consecutive order, the conductivity of the whole array was modified and thus the transmission dip frequency of the composites. The red-shifting of frequency when adding annealed microwires into the array mainly resulted from the decrease in internal damping due to annealing. Moreover, the results were used to forward predict the transmission dip frequency bands for a diversity of microwire arrays. Our single component-control approach provides a much simpler and lower cost platform than stimulus/hybridization-based methods for the design of microwave materials for smart wide frequency range applications.

**Acknowledgments**

This work was supported by NSFC No. 51671171, and No. 51811530103; RFBR No.18-58-53059/18; Basic Funding for Central Universities No. 2018QNA4001. Qin is also indebted to the support of the "National Youth Thousand Talent Program" of China.

**Supplementary data**

Surface morphology and topology of $Co_{60}Fe_{15}Si_{10}B_{15}$ glass-coated microwires, SEM and AFM analysis can be found in the supporting information.

**References**

[1]     Qin F, Brosseau C. A review and analysis of microwave absorption in polymer composites filled with carbonaceous particles. J Appl Phys 2012;111. doi:10.1063/1.3688435.

[2]     Qin F, Peng HX, Tang J, Qin LC. Ferromagnetic microwires enabled polymer composites for sensing applications. Compos Part A Appl Sci Manuf 2010;41:1823–8. doi:10.1016/j.compositesa.2010.09.003.

[3]     Luo Y, Peng HX, Qin FX, Ipatov M, Zhukova V, Zhukov A, et al. Metacomposite characteristics and their influential factors of polymer composites containing orthogonal ferromagnetic microwire arrays. J Appl Phys 2014;115:1–7. doi:10.1063/1.4874176.

[4]     Makhnovskiy D, Zhukov A, Zhukova V, Gonzalez J. Tunable and Self-Sensing

Microwave Composite Materials Incorporating Ferromagnetic Microwires. Adv Sci Technol 2008;54:201–10. doi:10.4028/www.scientific.net/AST.54.201.

[5]  Qin F, Peng HX. Ferromagnetic microwires enabled multifunctional composite materials. Prog Mater Sci 2013;58:183–259. doi:10.1016/j.pmatsci.2012.06.001.

[6]  Panina L V., Ipatov M, Zhukova V, Estevez J, Zhukov A. Microwave metamaterials containing magnetically soft microwires. Mater Res Soc Symp Proc 2011;1312:313–8. doi:10.1557/opl.2011.118.

[7]  Qin FX, Peng HX, Fuller J, Brosseau C. Magnetic field-dependent effective microwave properties of microwire-epoxy composites. Appl Phys Lett 2012;101. doi:10.1063/1.4758483.

[8]  Kostagiannakopoulou C, Tsilimigkra X, Sotiriadis G, Kostopoulos V. Synergy effect of carbon nano-fillers on the fracture toughness of structural composites. Compos Part B Eng 2017;129:18–25. doi:10.1016/j.compositesb.2017.07.012.

[9]  Mohamadi M, Kowsari E, Haddadi-Asl V, Yousefzadeh M. Fabrication, characterization and electromagnetic wave absorption properties of covalently modified reduced graphene oxide based on dinuclear cobalt complex. Compos Part B Eng 2019;162:569–79. doi:10.1016/j.compositesb.2019.01.032.

[10] Li Y, Wang S, Wang Q, Xing M. A comparison study on mechanical properties of polymer composites reinforced by carbon nanotubes and graphene sheet. Compos Part B Eng 2018;133:35–41. doi:10.1016/j.compositesb.2017.09.024.

[11] Ghorbanpour Arani A, Haghparast E, Khoddami Maraghi Z, Amir S. Static stress analysis of carbon nano-tube reinforced composite (CNTRC) cylinder under non-

axisymmetric thermo-mechanical loads and uniform electro-magnetic fields. Compos Part B Eng 2015;68:136–45. doi:10.1016/j.compositesb.2014.08.036.

[12] Estevez D, Qin FX, Quan L, Luo Y, Zheng XF, Wang H, et al. Complementary design of nano-carbon/magnetic microwire hybrid fibers for tunable microwave absorption. Carbon N Y 2018;132:486–94. doi:10.1016/j.carbon.2018.02.083.

[13] Estevez D, Peng H-X, Quan L, Mai Y-W, Qin F, Panina L, et al. Tunable negative permittivity in nano-carbon coated magnetic microwire polymer metacomposites. Compos Sci Technol 2018;171:206–17. doi:10.1016/j.compscitech.2018.12.016.

[14] Larin VS, Torcunov A V, Zhukov A, Vazquez M, Panina L. Preparation and properties of glass-coated microwires 2002;249:39–45.

[15] Vinod EM, Ramesh K, Sangunni KS. Structural transition and enhanced phase transition properties of Se doped Ge2Sb2Te5 alloys. Sci Rep 2015;5:1–7. doi:10.1038/srep08050.

[16] Marks RB, Jargon JA, Juroshek JR. Calibration Comparison Method for Vector Network Analyzers. October 1996:38–45. doi:10.1109/ARFTG.1996.327186.

[17] Dzhumazoda A, Panina L V., Nematov MG, Ukhasov AA, Yudanov NA, Morchenko AT, et al. Temperature-stable magnetoimpedance (MI) of current-annealed Co-based amorphous microwires. J Magn Magn Mater 2019;474:374–80. doi:10.1016/j.jmmm.2018.10.111.

[18] Nematov MG, Adam AM, Panina L V., Yudanov NA, Dzhumazoda A, Morchenko AT, et al. Magnetic anisotropy and stress-magnetoimpedance (S-MI)

in current-annealed Co-rich glass-coated microwires with positive magnetostriction. J Magn Magn Mater 2019;474:296–300. doi:10.1016/j.jmmm.2018.11.042.

[19]  Sergey Kaloshkin, Margarita Churyukanova VT. Characterization of Magnetic Transformation at Curie Temperature in Finemet-type Microwires 2012;1408:1–6. doi:10.1557/opl.201.

[20]  Zhukova V, Kaloshkin S, Zhukov A, Gonzalez J. DSC studies of finemet-type glass-coated microwires. J Magn Magn Mater 2002;249:108–12. doi:10.1016/S0304-8853(02)00515-2.

[21]  Eckert J, Mattern N, Kim WT, Kim DH, Park ES. Phase separation in metallic glasses. Prog Mater Sci 2013;58:1103–72. doi:10.1016/j.pmatsci.2013.04.002.

[22]  Uddin A, Evstigneeva SA, Dzhumazoda A, Salem MM, Nematov MG, Adam AM, et al. Temperature Effects on the Magnetization and Magnetoimpedance in Ferromagnetic Glass-Covered microwires. J. Phys. Conf. Ser., 2017. doi:10.1088/1742-6596/917/8/082011.

[23]  Vázquez M, Chiriac H, Zhukov A, Panina L, Uchiyama T. On the state-of-the-art in magnetic microwires and expected trends for scientific and technological studies. Phys Status Solidi Appl Mater Sci 2011;208:493–501. doi:10.1002/pssa.201026488.

[24]  Jiang SD, Eggers T, Thiabgoh O, Xing DW, Fei WD, Shen HX, et al. Relating surface roughness and magnetic domain structure to giant magneto-impedance of Co-rich melt-extracted microwires. Sci Rep 2017;7:1–8. doi:10.1038/srep46253.

[25]  Melo LGC, Ménard D, Ciureanu P, Yelon A. Influence of surface anisotropy on magnetoimpedance in wires. J Appl Phys 2002;92:7272–80. doi:10.1063/1.1519345.

[26]  Wang H, Qin FX, Xing DW, Cao FY, Wang XD, Peng HX, et al. Relating residual stress and microstructure to mechanical and giant magneto-impedance properties in cold-drawn Co-based amorphous microwires. Acta Mater 2012;60:5425–36. doi:10.1016/j.actamat.2012.06.047.

[27]  Kraus L. GMI modeling and material optimization. Sensors Actuators, A Phys 2003;106:187–94. doi:10.1016/S0924-4247(03)00164-X.

[28]  Chiriac H, Óvári TA, Zhukov A. Magnetoelastic anisotropy of amorphous microwires. J Magn Magn Mater 2003;254–255:469–71. doi:10.1016/S0304-8853(02)00875-2.

[29]  Vázquez M, Pirota K, Torrejón J, Badini G, Torcunov A. Magnetoelastic interactions in multilayer microwires. J Magn Magn Mater 2006;304:197–202. doi:10.1016/j.jmmm.2006.02.122.

[30]  Qin FX, Tang J, Popov V V., Liu JS, Peng HX, Brosseau C. Influence of direct bias current on the electromagnetic properties of melt-extracted microwires and their composites. Appl Phys Lett 2014;104:1–5. doi:10.1063/1.4861185.

[31]  Jiang S, Xing D, Liu J, Shen H, Chen D, Fang W, et al. Influence of microstructure evolution on GMI properties and magnetic domains of melt-extracted Zr-doped amorphous wires with accumulated DC annealing. J Alloys Compd 2015;644:180–5. doi:10.1016/j.jallcom.2015.04.187.

[32] Zhukov A, Ipatov M, del Val JJ, Churyukanova M, Zhukova V. Tailoring of magnetic properties of Heusler-type glass-coated microwires by annealing. J Alloys Compd 2018;732:561–6. doi:10.1016/j.jallcom.2017.10.232.

[33] Klein P, Varga R, Onufer J, Ziman J, Badini-Confalonieri GA, Vazquez M. Effect of current annealing on domain structure in amorphous and nanocrystalline FeCoMoB microwires. Acta Phys Pol A 2017;131:681–3. doi:10.12693/APhysPolA.131.681.

[34] Kildishev A V., Shalaev VM, Chen K-P, Borneman JD, Drachev VP. Drude Relaxation Rate in Grained Gold Nanoantennas. Nano Lett 2010;10:916–22. doi:10.1021/nl9037246.

[35] Mosbah A, Aida MS. Influence of deposition temperature on structural, optical and electrical properties of sputtered Al doped ZnO thin films. J Alloys Compd 2012;515:149–53. doi:10.1016/j.jallcom.2011.11.113.

[36] V. P. Deshpande, S. D. Sartale, A. N. Vyas AUU. Temperature Dependent Properties of Spray Deposited Nanostructured ZnO Thin Films. Int J Mater Chem 2017;7:36–46. doi:10.5923/j.ijmc.20170702.03.

**Figure Captions:**

**Figure 1**: Schematic of current annealing of $Co_{60}Fe_{15}Si_{10}B_{15}$ microwire by applying DC current.

**Figure 2:** Schematic of the arrangement of the wires; (a) wire composites incorporating the individual as-cast wires: A, 30 mA: B and 40 mA: C; (b) and(c) correspond to their combination and different location in the composite, respectively.

Figure 3: (a) Differential Scanning Calorimetry (DSC) of as-cast $Co_{60}Fe_{15}Si_{10}B_{15}$ microwires, where $T_c$ and $T_x$ correspond to the Curie and crystallization temperature respectively; The insert shows the inflexion point on the DSC curve for as-cast wire which was used for the determination of $T_c$. (b) XRD spectra for as-cast $Co_{60}Fe_{15}Si_{10}B_{15}$ microwires (black spectrum) and current annealed wires at 40 mA (red spectrum), respectively.

**Figure 4:** HRTEM images of as-cast microwire (a) and microwire subjected to thermal annealing of 30 mA (b) and 40 mA (c). The insets show the SAED pattern for the as-cast wire and the enlarged view of HRTEM images for 30 and 40 mA-annealed samples to highlight the droplet and nanocrystalline structure. "A" corresponds to the amorphous phase. The yellow arrows indicate the bright outline around the droplet composed mainly by the lighter elements Si and B.

**Figure 5**: HAADF TEM images of $Co_{60}Fe_{15}Si_{10}B_{15}$ (a) as-cast microwires and the corresponding EDS mapping for (a1) Co (a2) Fe and (a3) Si respectively.

**Figure 6:** (a) Magnetic hysteresis loops of the $Co_{60}Fe_{15}Si_{10}B_{15}$ microwires of as-cast and current annealed microwires at 20 mA, 30 mA and 40 mA. Insets show the zoom-in M-H loops at low field and the magnetic domain structure of the as-cast and 40mA-annealed

wire. (b) Real part of impedance at 200 MHz for as-cast and current annealed wires at 30 mA and 40 mA.

**Figure 7:** (a); (b) and (c) Transmission spectra $S_{21}$ for composites containing the individual as-cast and annealed microwires and the composites containing different combinations of the as-cast, 30 and 40 mA-annealed microwires, respectively.

**Figure 8:** Interaction between the wires with different arrangements in the composite and their corresponding "effective area", (a) AAAXXX (b) AXXAAX (c) AXAXAX, where X corresponding to either 30 or 40 mA annealed microwire. Wire-wire interactions and couplings are stronger when the wires are in the vicinity of their twins.

**Figure 9:** Transmission dip frequency shift in response to different array arrangements in the composite, where A and X correspond to the as-cast wire and annealed wires, respectively.

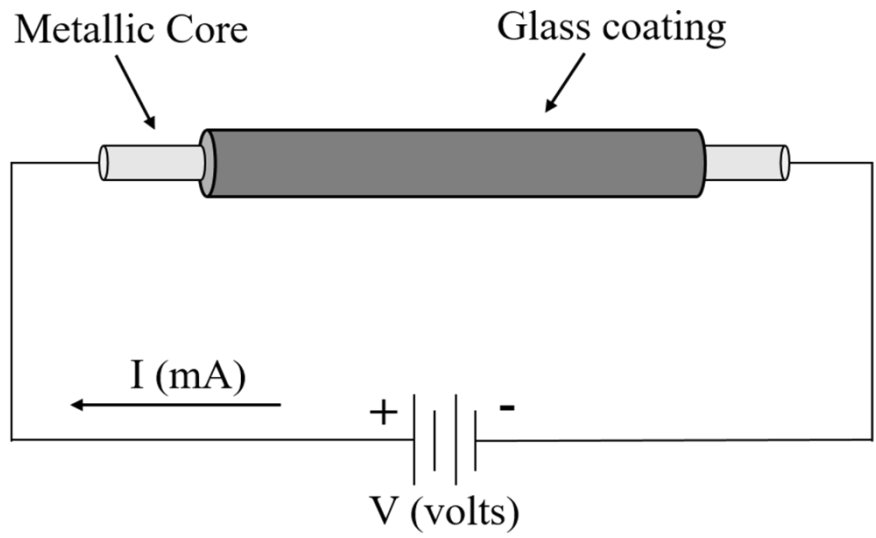

**Figure 1**: Schematic of current annealing of $Co_{60}Fe_{15}Si_{10}B_{15}$ microwire by applying DC current

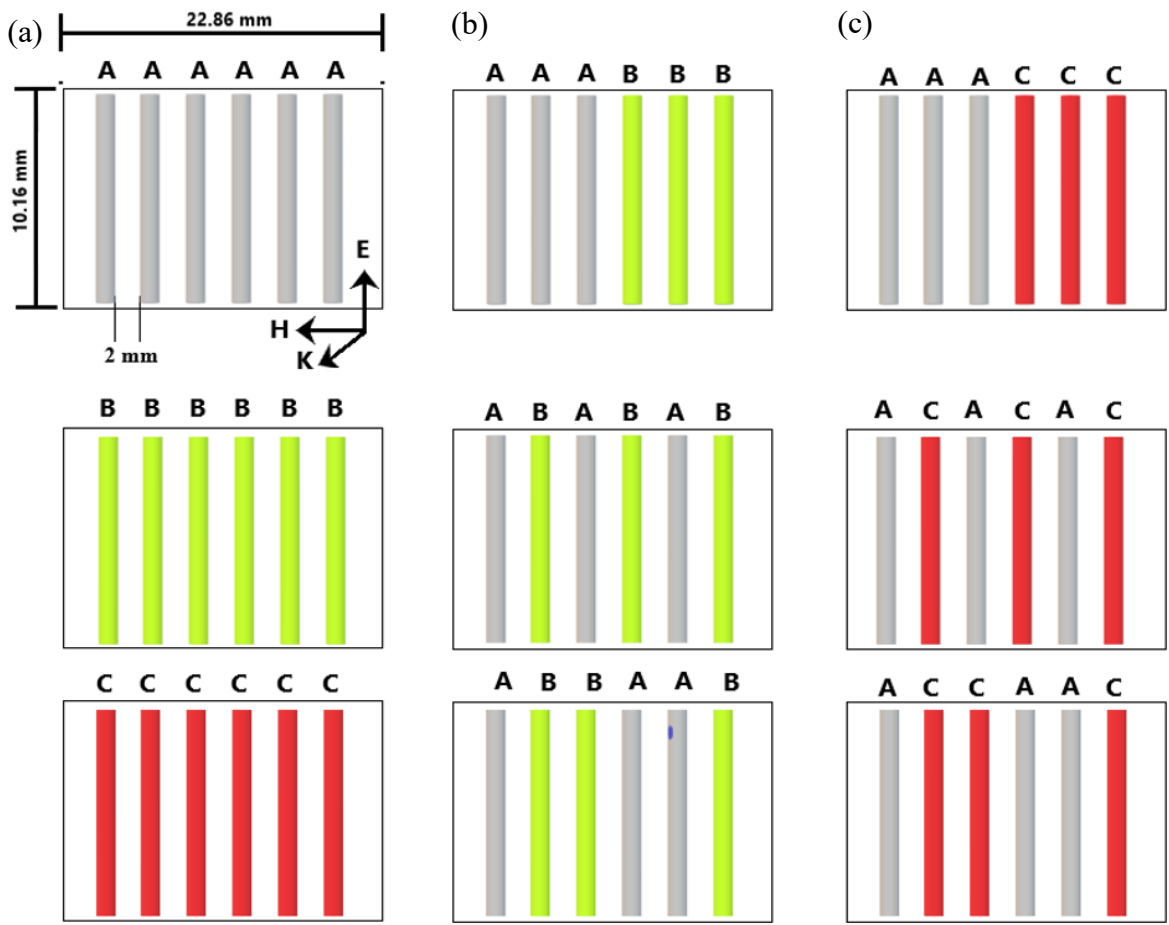

**Figure 2:** Schematic of the arrangement of the wires; (a) wire composites incorporating the individual as-cast wires: A, 30 mA: B and 40 mA: C; (b) and(c) correspond to their combination and different location in the composite, respectively.

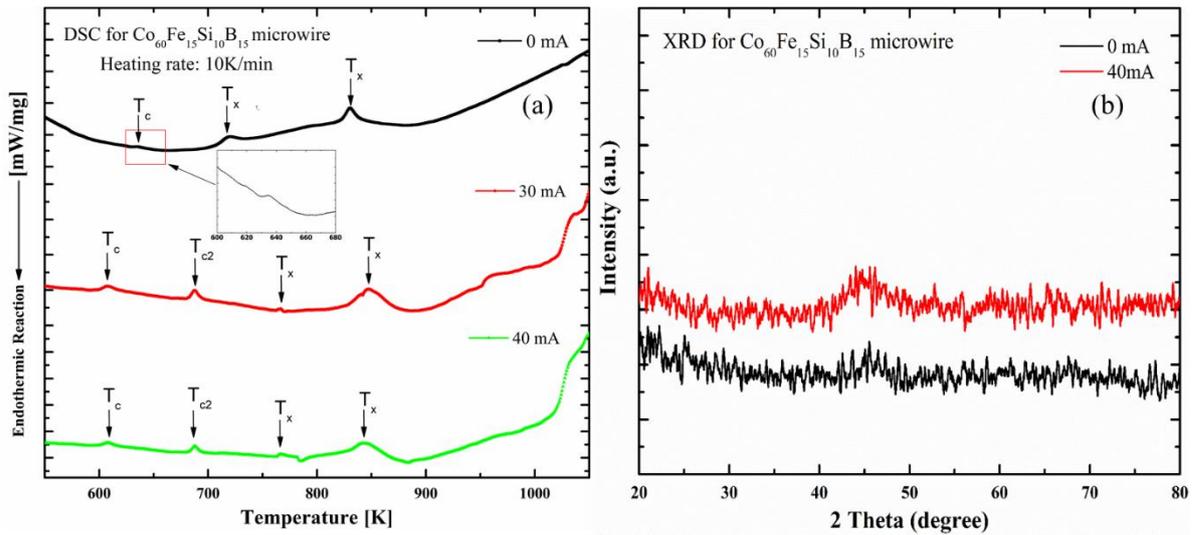

Figure 3: (a) Differential Scanning Calorimetry (DSC) of as-cast $Co_{60}Fe_{15}Si_{10}B_{15}$ microwires, where $T_c$ and $T_x$ correspond to the Curie and crystallization temperature respectively; The insert shows the inflexion point on the DSC curve for as-cast wire which was used for the determination of $T_c$. (b) XRD spectra for as-cast $Co_{60}Fe_{15}Si_{10}B_{15}$ microwires (black spectrum) and current annealed wires at 40 mA (red spectrum), respectively.

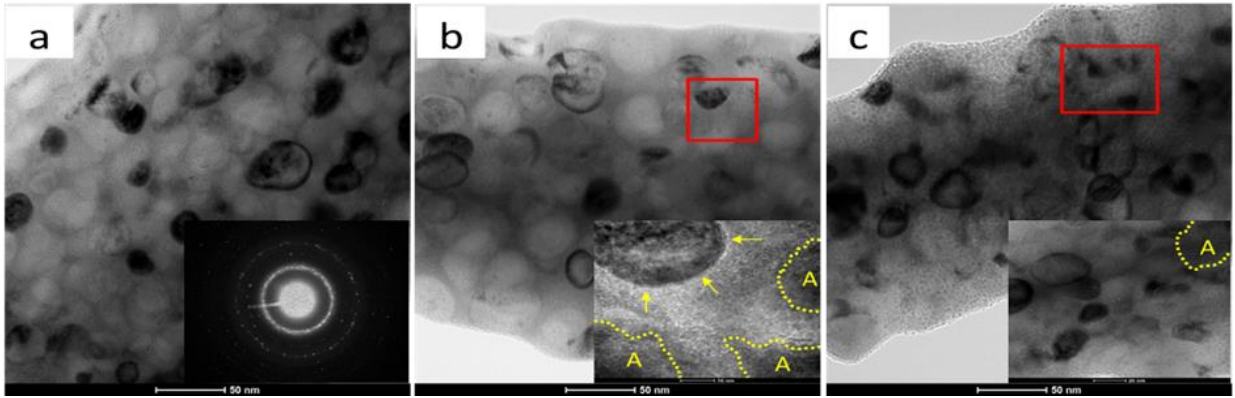

**Figure 4:** HRTEM images of as-cast microwire (a) and microwire subjected to thermal annealing of 30 mA (b) and 40 mA (c). The insets show the SAED pattern for the as-cast wire and the enlarged view of HRTEM images for 30 and 40 mA-annealed samples to highlight the droplet and nanocrystalline structure. "A" corresponds to the amorphous phase. The yellow arrows indicate the bright outline around the droplet composed mainly by the lighter elements Si and B.

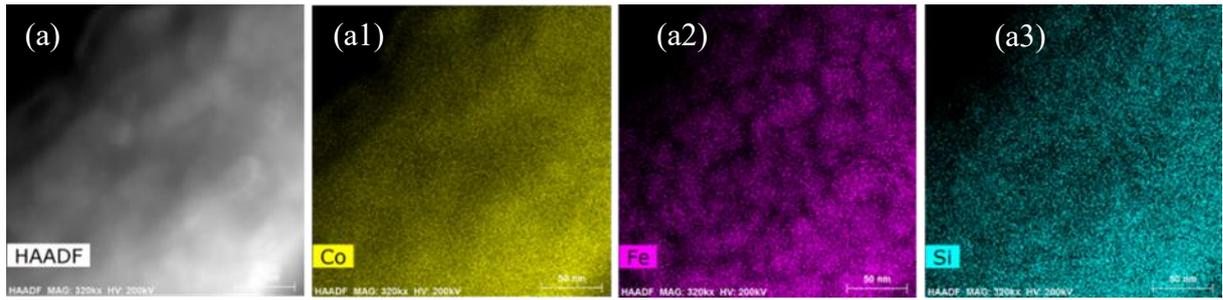

**Figure 5**: HAADF TEM images of $Co_{60}Fe_{15}Si_{10}B_{15}$ (a) as-cast microwires and the corresponding EDS mapping for (a1) Co (a2) Fe and (a3) Si respectively.

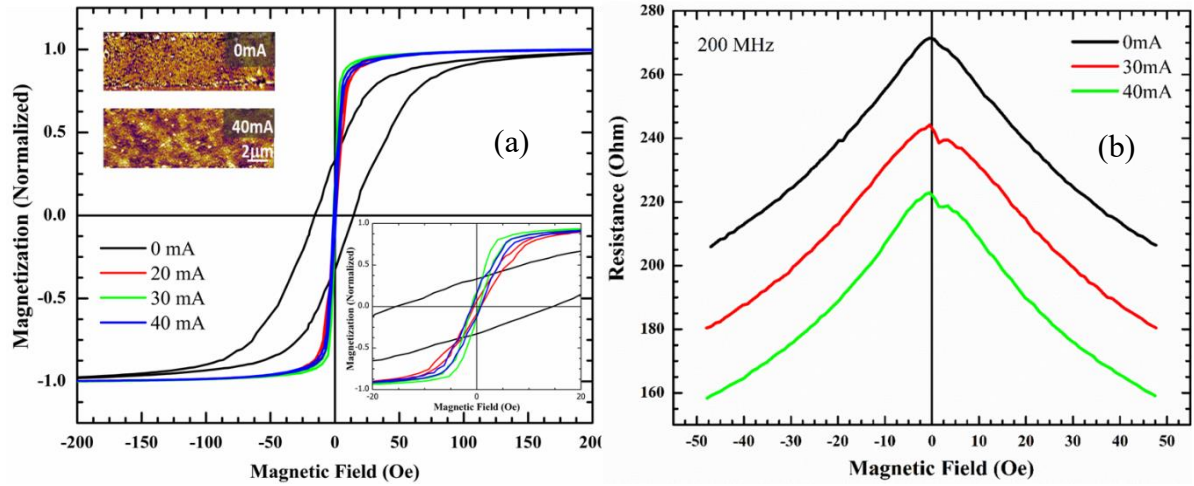

**Figure 6:** (a) Magnetic hysteresis loops of the $Co_{60}Fe_{15}Si_{10}B_{15}$ microwires of as-cast and current annealed microwires at 20 mA, 30 mA and 40 mA. Insets show the zoom-in M-H loops at low field and the magnetic domain structure of the as-cast and 40mA-annealed wire. (b) Real part of impedance at 200 MHz for as-cast and current annealed wires at 30 mA and 40 mA.

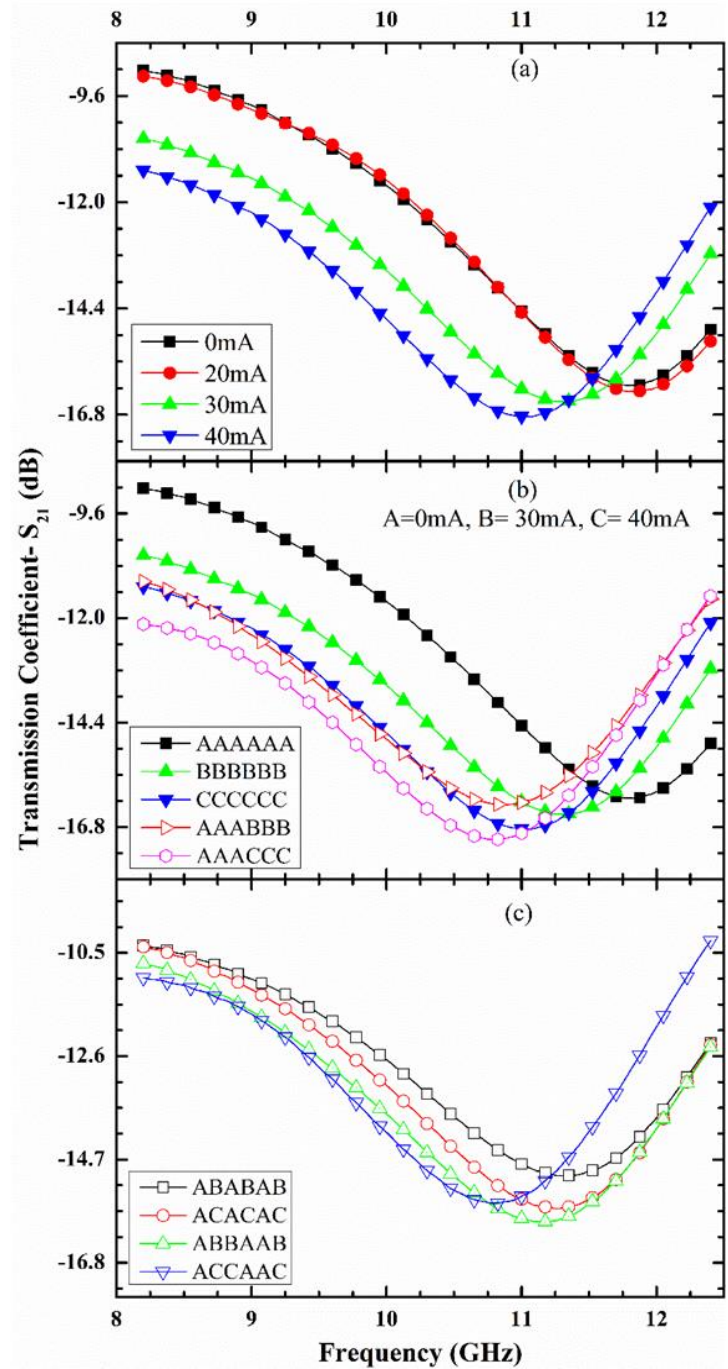

**Figure 7:** (a); (b) and (c) Transmission spectra $S_{21}$ for composites containing the individual as-cast and annealed microwires and the composites containing different combinations of the as-cast, 30 and 40 mA-annealed microwires, respectively.

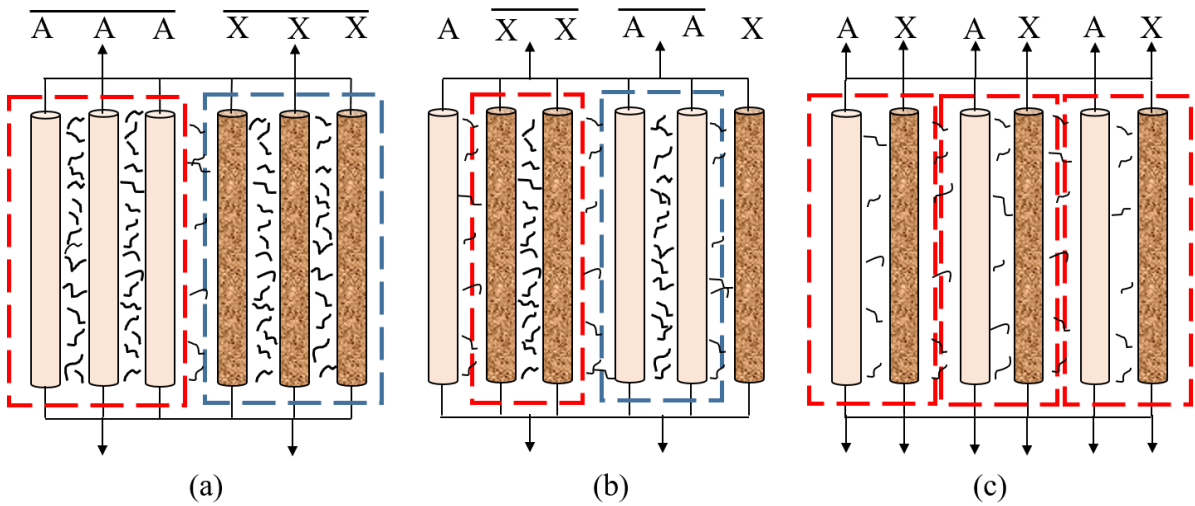

**Figure 8:** Interaction between the wires with different arrangements in the composite and their corresponding "effective area", (a) AAAXXX (b) AXXAAX (c) AXAXAX, where X corresponding to either 30 or 40 mA annealed microwire. Wire-wire interactions and couplings are stronger when the wires are in the vicinity of their twins.

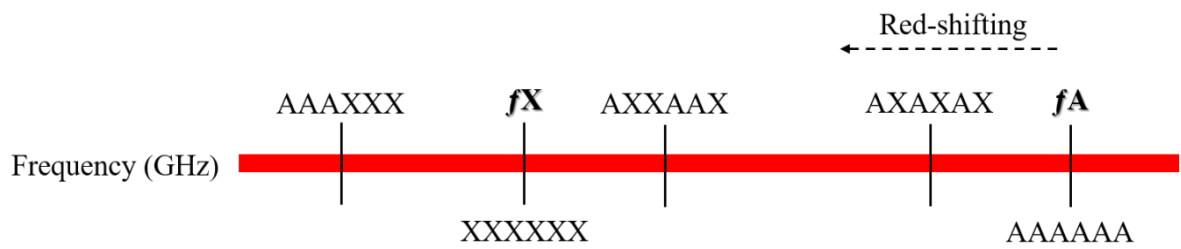

**Figure 9:** Transmission dip frequency shift in response to different array arrangements in the composite, where A and X correspond to the as-cast wire and annealed wires, respectively.